\documentclass[prl, superscriptaddress, twocolumn, 10pt,showpacs]{revtex4}  
\usepackage{amssymb}
\usepackage{amsmath}
\usepackage{bbm}
\usepackage{graphicx}
\usepackage{color}

\begin{document}

\author{Philipp Werner}
\affiliation{Department of Physics, University of Fribourg, 1700 Fribourg, Switzerland}
\author{Xi Chen}
\affiliation{Center for Computational Quantum Physics, Flatiron Institute, 162 Fifth Avenue, New York, NY 10010, USA}
\author{Emanuel Gull}
\affiliation{Department of Physics, University of Michigan, Ann Arbor, MI 48109, USA}

\title{Ferromagnetic spin correlations in the two-dimensional Hubbard model} 

\begin{abstract}
We analyze the dynamical nearest-neighbor and next-nearest-neighbor spin correlations in the 4-site and 8-site dynamical cluster approximation to the two-dimensional Hubbard model. Focusing on the robustness of these correlations at long imaginary times, we reveal enhanced ferromagnetic correlations on the lattice diagonal, consistent with the emergence of composite spin-1 moments at a temperature scale that essentially coincides with the pseudo-gap temperature $T^*$. We discuss these results in the context of the spin-freezing theory of unconventional superconductivity.
\end{abstract}

\date{\today}

\hyphenation{}


\maketitle

The two-dimensional (2D) Hubbard model has been studied extensively in connection with high-temperature superconductivity in the cuprates \cite{Anderson1987,Dagotto1994,Scalapino2007,Zheng2017}. It is believed to be the simplest model that captures the relevant physics in these compounds and indeed, many features of the cuprate phase diagram can be reproduced using, for example, cluster extensions of dynamical mean field theory (DMFT) \cite{Lichtenstein1999, Jarrell2000, Kotliar2001, Maier2005}. These studies, most of them based on a 4-site plaquette embedded in a self-consistent dynamical mean field, not only reproduced a $d$-wave superconducting dome next to or partially overlapping with an antiferromagnetic phase \cite{Lichtenstein1999,Jarrell2000,Kancharla2008}, but also the characteristic signatures of the pseudo-gap state \cite{Civelli2005,Kyung2006,Gull2009}. They further revealed that the pseudo-gap state competes with superconductivity \cite{Gull2013}. 

Despite this progress, the actual mechanism underlying the appearance of the pseudo-gap and high-temperature superconductivity remains to be understood. A common view is that antiferromagnetic spin fluctuations provide the ``glue" for superconductivity \cite{Scalapino2012}. The recently proposed spin-freezing theory of unconventional superconductity \cite{Hoshino2015,Werner2016} on the other hand suggests that fluctuations in the magnitude of a composite spin (henceforth referred to as local spin fluctuations) play a crucial role. This motivates the present study of dynamical spin correlations in the 2D Hubbard model.

\begin{figure}[t]
\begin{center}
\includegraphics[angle=0, width=0.8\columnwidth]{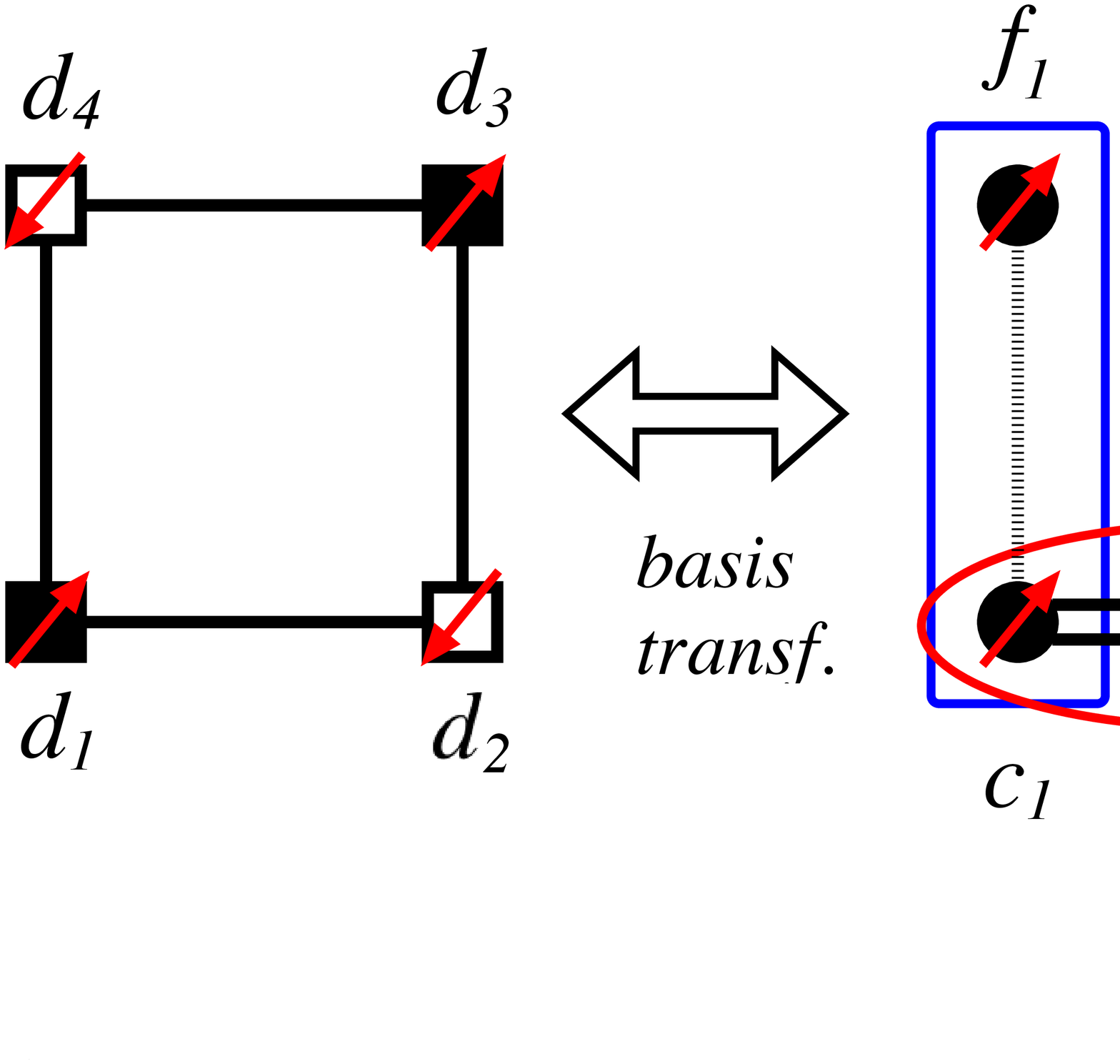}
\caption{
Illustration of the basis transformation from the 4-site single-orbital plaquette (left) to a 2-site two-orbital cluster (right). Solid black lines represent a hopping $t$, double lines a hopping $2t$ and dashed lines a Slater-Kanamori interaction with $\tilde U=\tilde U'=\tilde J=U/2$ (where $U$ is the Hubbard interaction on the plaquette). Blue rectangles represent the tendency to spin-1 formation on a given site of the 2-orbital cluster, and the red ellipse the competing tendency to form a spin-$\tfrac 12$ singlet between $c$ electrons with opposite spins. 
}
\label{fig_illustration}
\end{center}
\end{figure}

The link to spin-freezing induced superconductivity \cite{Hoshino2015} and Hund metal behavior \cite{Georges2013} is provided by the 4-site cluster DMFT construction. This cluster DMFT solution can be mapped exactly to an auxiliary 2-site, two-orbital cluster problem with a Slater-Kanamori interaction on each site \cite{Werner2016}. The mapping involves a transformation to bonding and antibonding orbitals along the diagonals of the 4-site cluster, as illustrated in Fig.~\ref{fig_illustration}. If the annihilation operators on the 4-site cluster are denoted by $d_1,\ldots, d_4$ the transformed orbitals are $f_1=\frac{1}{\sqrt{2}}(d_1-d_3)$, $c_1=\frac{1}{\sqrt{2}}(d_1+d_3)$, $f_2=\frac{1}{\sqrt{2}}(d_2-d_4)$, $c_2=\frac{1}{\sqrt{2}}(d_2+d_4)$. The freezing of spins (due to the ferromagnetic Hund coupling, see blue rectangles in Fig.~\ref{fig_illustration}) in the effective two-orbital description translates into robust composite spin-1 moments on the diagonals of the original cluster. An interesting open question concerns the antiferromagnetic correlations between these spin-1 moments, and the competition with spin-$\tfrac12$ nearest-neighbor singlet formation (red ellipse), i.e., the antiferromangetic correlations that have been considered in most of the previous literature \cite{ellipse}. It is also relevant to ask how robust this picture is as one moves to larger clusters.  

The spin freezing theory focuses on cluster spin correlations measured in the bonding and antibonding basis, such as 
$S_{ff}(\tau)=\langle S^z_f(\tau) S^z_f(0)\rangle$. In terms of the original operators, and for $f\equiv f_1$, we can write $S^z_f = \tfrac{1}{2}[ S^z_1 + S^z_3 - \tfrac{1}{2}(Y_\uparrow - Y_\downarrow)]$ with $Y_\sigma=d^\dagger_{1\sigma}d_{3\sigma} + d^\dagger_{3\sigma}d_{1\sigma}$. The $S_{ff}$ correlator can thus be expressed as $S_{ff}(\tau)=\tfrac{1}{2}[S_{11}(\tau)+S_{13}(\tau)]-\tfrac{1}{8}[\langle (S^z_1(0)+S^z_3(0))(Y_\uparrow(\tau) - Y_\downarrow(\tau))\rangle]-\tfrac{1}{8}[\langle (Y_\uparrow(0) - Y_\downarrow(0)) (S^z_1(\tau)+S^z_3(\tau))\rangle]$, and requires a measurement of correlations functions of the type $\langle n_{1\sigma}(\tau) d^\dagger_{1\sigma'}d_{3\sigma'}(0)\rangle$. This is numerically challenging and requires a worm-type sampling \cite{Gunacker2015,Werner2016}. Some results for the easily measurable contribution $\tfrac{1}{2}[S_{11}(\tau)+S_{13}(\tau)]$ are presented in the Supplementary Material.

In the present study, we take a step back and instead of discussing the dynamics of $S_{ff}(\tau)$ compute and analyze the nearest-neighbor $S_{12}$ and diagonal next-nearest neighbor $S_{13}$ spin correlations in the original basis. We employ the dynamical cluster approximation (DCA) \cite{Hettler1998}, which enforces translational invariance on the cluster, and report the spin correlations measured on the impurity cluster. The calculations are performed for a square lattice Hubbard model with nearest-neighbor hopping $t$, which we use as the unit of energy. The on-site repulsion is fixed to $U=8$. 

\begin{figure}[t]
\begin{center}
\includegraphics[angle=-90, width=0.49\columnwidth]{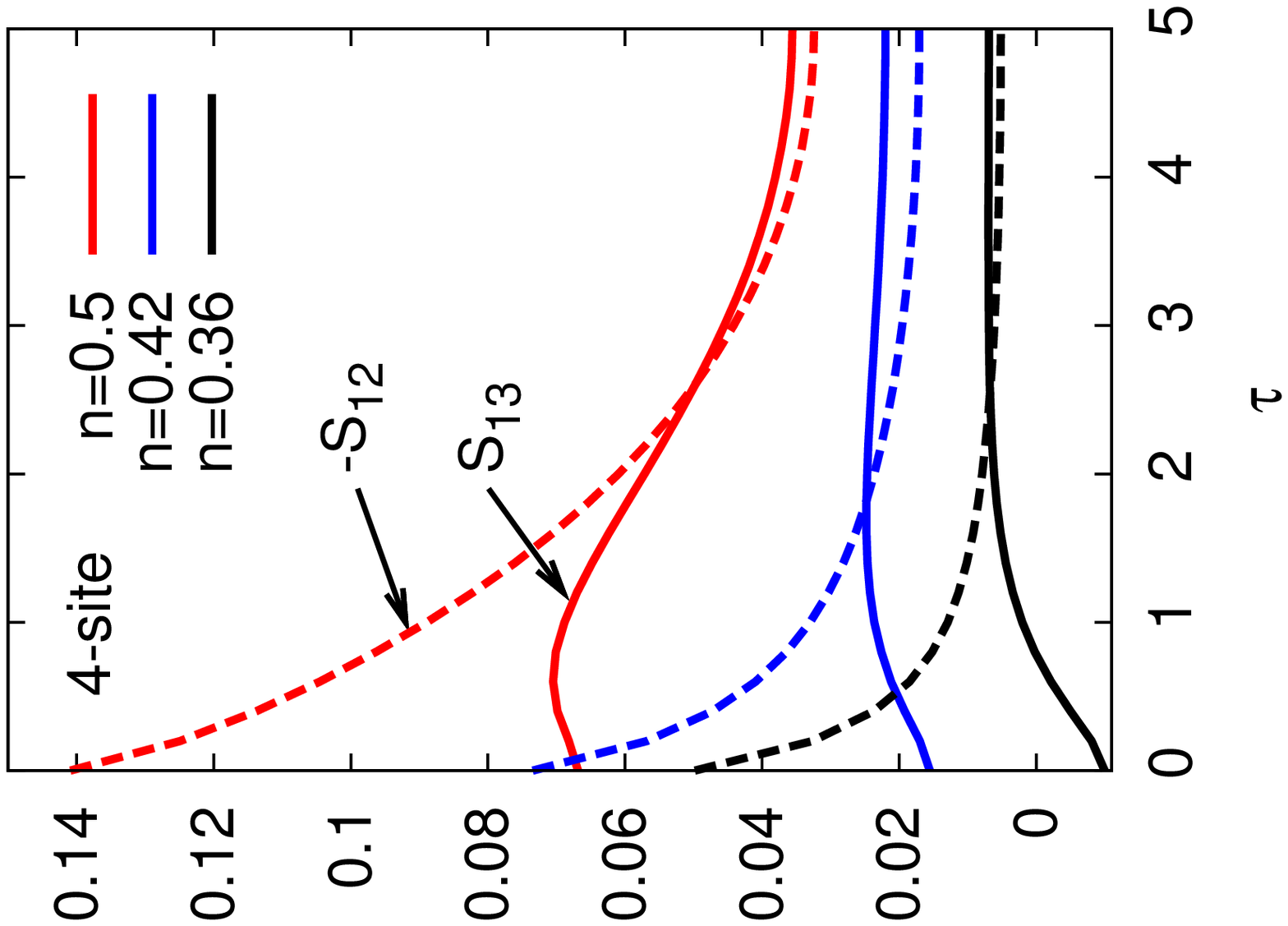}\hfill
\includegraphics[angle=-90, width=0.49\columnwidth]{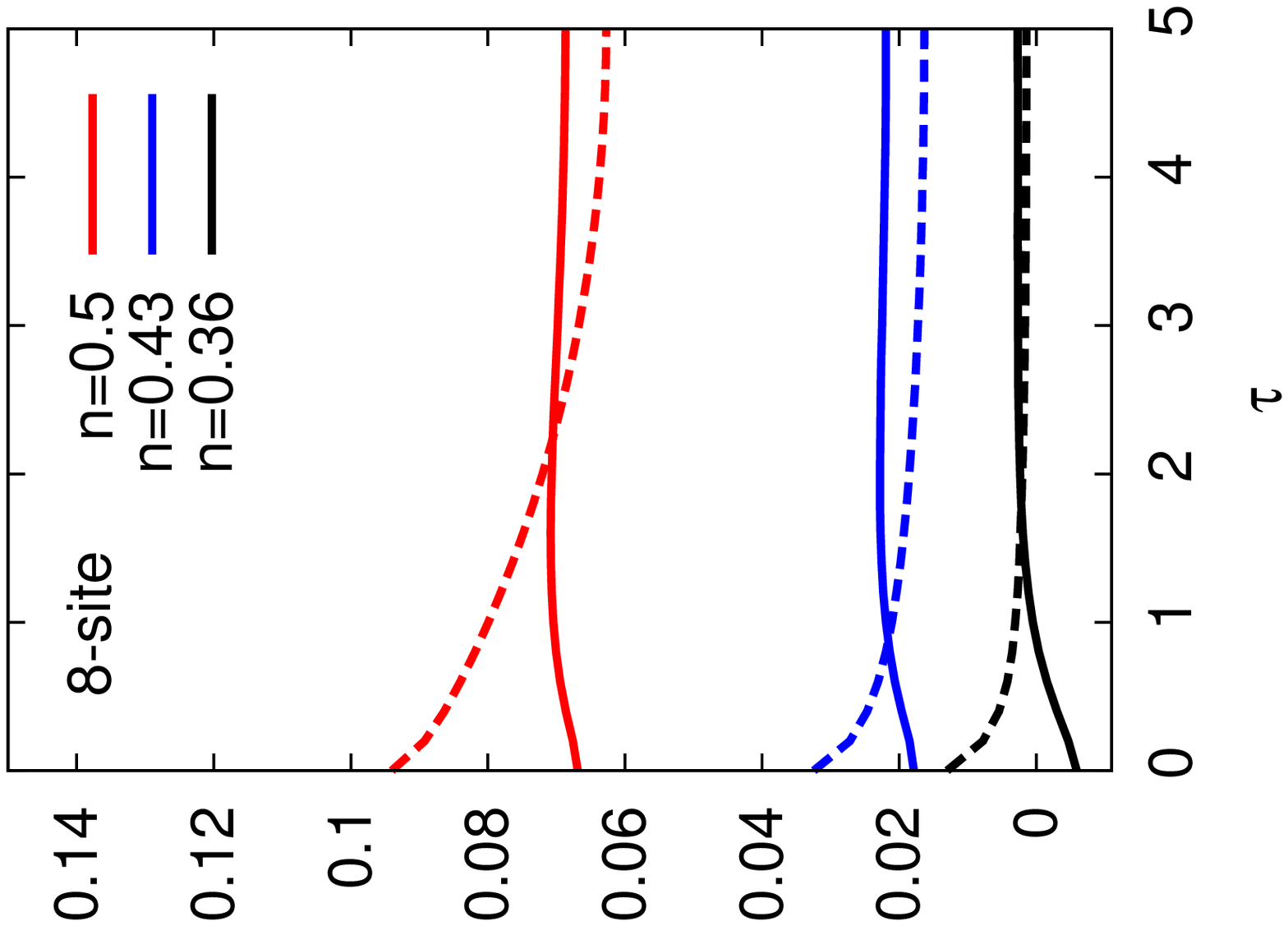}
\caption{
Imaginary-time dependent nearest-neighbor ($-S_{12}$, dashed) and diagonal next-nearest-neighbor ($S_{13}$, solid) spin correlations at $\beta=10$ and indicated fillings, in 4-site DCA (left) and 8-site DCA (right). While antiferromagnetic $S_{12}$ correlations dominate at short time, they are weaker than the ferromagnetic $S_{13}$ correlations at $\tau=\beta/2$.
}
\label{fig_corr}
\end{center}
\end{figure}

Figure~\ref{fig_corr} plots $-S_{12}$ and $S_{13}$ for inverse temperature $\beta=10$ and different fillings (half filling corresponds to $n_\sigma=0.5$). The left (right) panel reports results for the 4-site (8-site) cluster. At short imaginary times $\tau$ the antiferromagnetic nearest-neighbor correlations (dashed lines) dominate the diagonal next-nearest-neighbor correlations (solid lines). The time dependence is however nontrivial, and for a broad range of fillings, we find that at long times the ferromangetic $S_{13}$ correlations dominate. This is related to the interesting fact that $S_{13}$ can increase with increasing $\tau$, in contrast to $-S_{12}$, which always decreases. We interpret the robustness of $S_{13}$ as a signature of the formation of a composite spin-1 moment on diagonally opposite sites. 
While this phenomenon has been anticipated in the previous spin-freezing analysis based on 4-site cluster DMFT \cite{Werner2016}, it is manifest also in the DCA solution and persists in the 8-site calculation.  

\begin{figure}[t]
\begin{center}
\includegraphics[angle=-90, width=0.9\columnwidth]{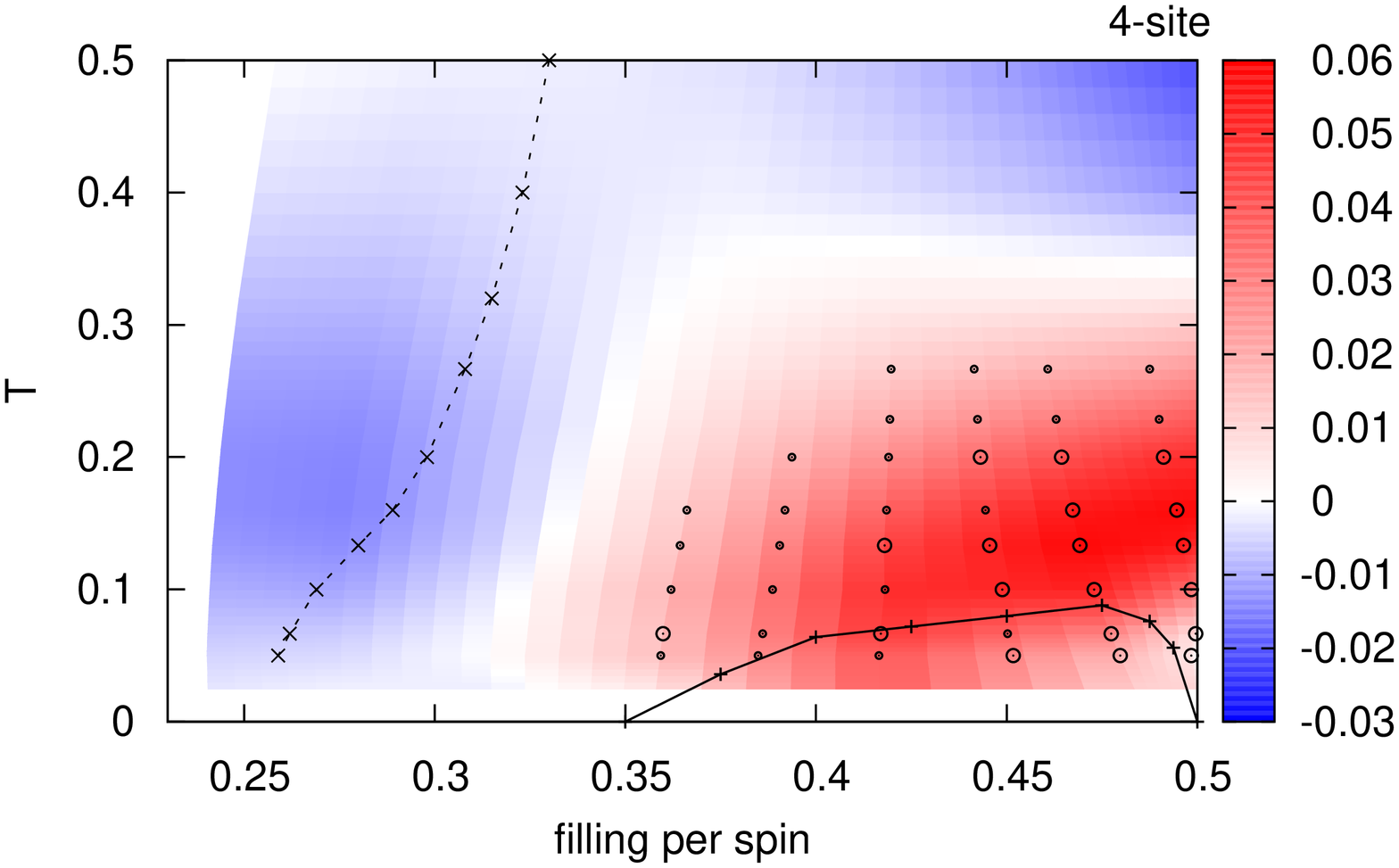} 
\includegraphics[angle=-90, width=0.9\columnwidth]{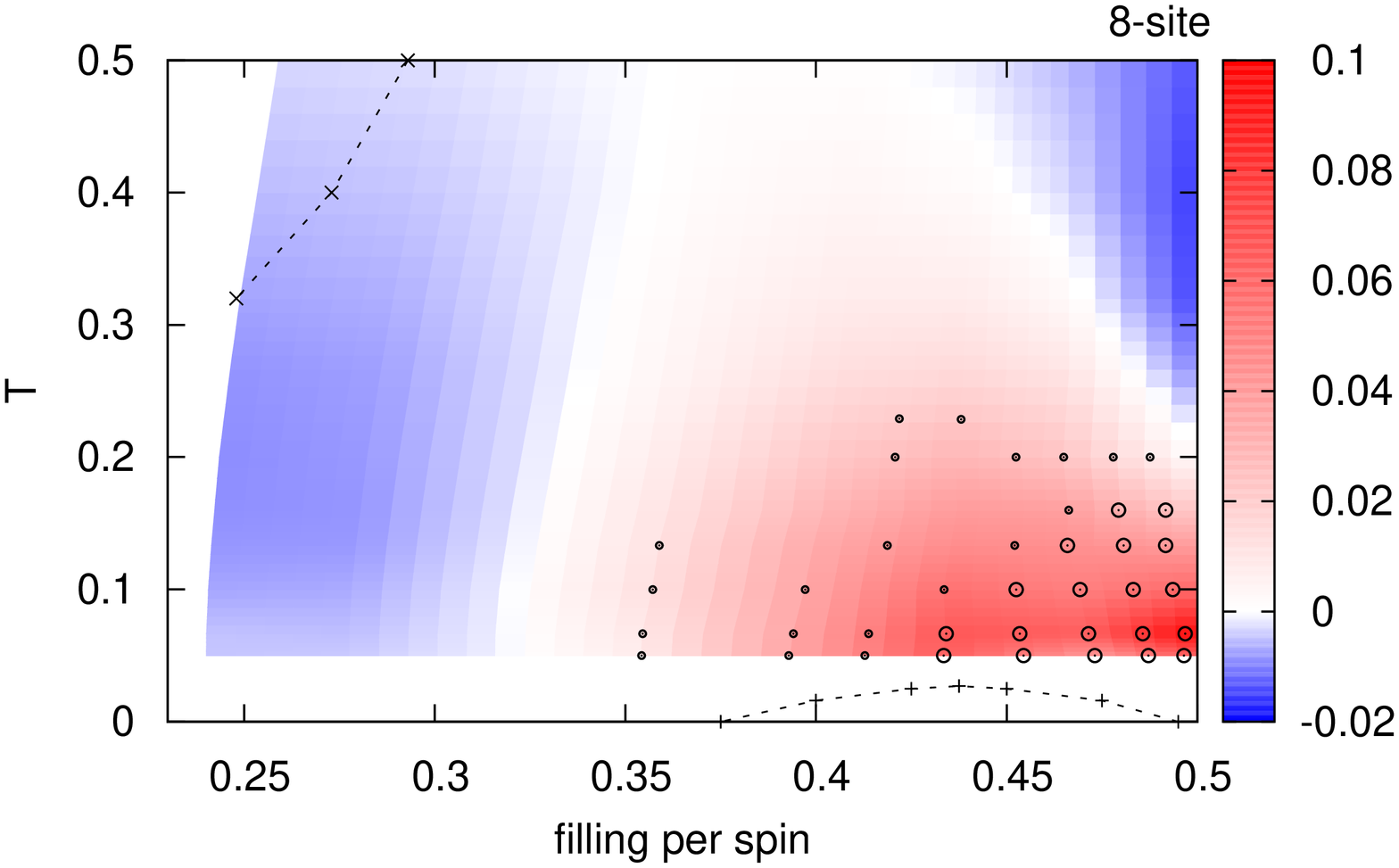} 
\caption{
Difference in spin correlations $\beta[S_{13}(\beta/2)-(-S_{12}(\beta/2))]$ in 4-site DCA (top) and 8-site DCA (bottom). The $T_c$ data for the $d$-wave superconducting phase in the 4-site case (solid black line) are taken from Ref.~\onlinecite{Jarrell2000}. The $T_c$ dome of the 8-site cluster is an educated guess based on the results in Ref.~\onlinecite{Dong2019}. Dashed lines with diagonal crosses indicate the filling below which the nearest-neighbor spin correlations at $\tau=\beta/2$ become ferromagnetic.  
}
\label{fig_corrtimesbeta}
\end{center}
\end{figure}

Because the long-time behavior reveals interesting properties of the spin correlations, we will from now on focus on the values of $S_{12}$ and $S_{13}$ at $\tau=\beta/2$. In Fig.~\ref{fig_corrtimesbeta} we plot $\beta[S_{13}(\beta/2)-(-S_{12}(\beta/2))]$, where the multiplication with $\beta$ is meant to compensate for the overall decay of the correlations with time, but is not crucial for the following analysis. A positive value indicates dominant ferromagnetic correlations on the diagonals, while a negative value implies dominant antiferromagnetic nearest-neighbor correlations. Let us first focus on the 4-site DCA results, shown in the top panel. Here, we also indicate by a black solid line the $T_c$ dome previously computed for the same model in Ref.~\onlinecite{Jarrell2000}. It is apparent from this plot that ferromagnetic next-nearest neighbor correlations dominate over a wide doping range, and in fact almost exactly the doping range in which $d$-wave superconductivity is found at low temperatures. 

The strongest enhancement of ferromagnetic correlations occurs in a temperature and doping region that one typically associates with the onset of the pseudo-gap (``$T^*$ line"). This suggests that the formation and freezing of composite spins on the lattice diagonals is at the root of the pseudo-gap phenomenon. Note that in our simulations, long-range order is suppressed. It is natural to assume that the spin-1 moments formed on the diagonals will order antiferromagnetically close to half-filling. In fact, the region with dominant ferromagnetic correlations is also quite similar to the antiferromagnetic region reported in Ref.~\onlinecite{Jarrell2000}. Whether or not the composite spin-1 moments will still be present in the long-range ordered antiferromagnetic phase is an interesting open question. 

\begin{figure}[t]
\begin{center}
\includegraphics[angle=-90, width=0.49\columnwidth]{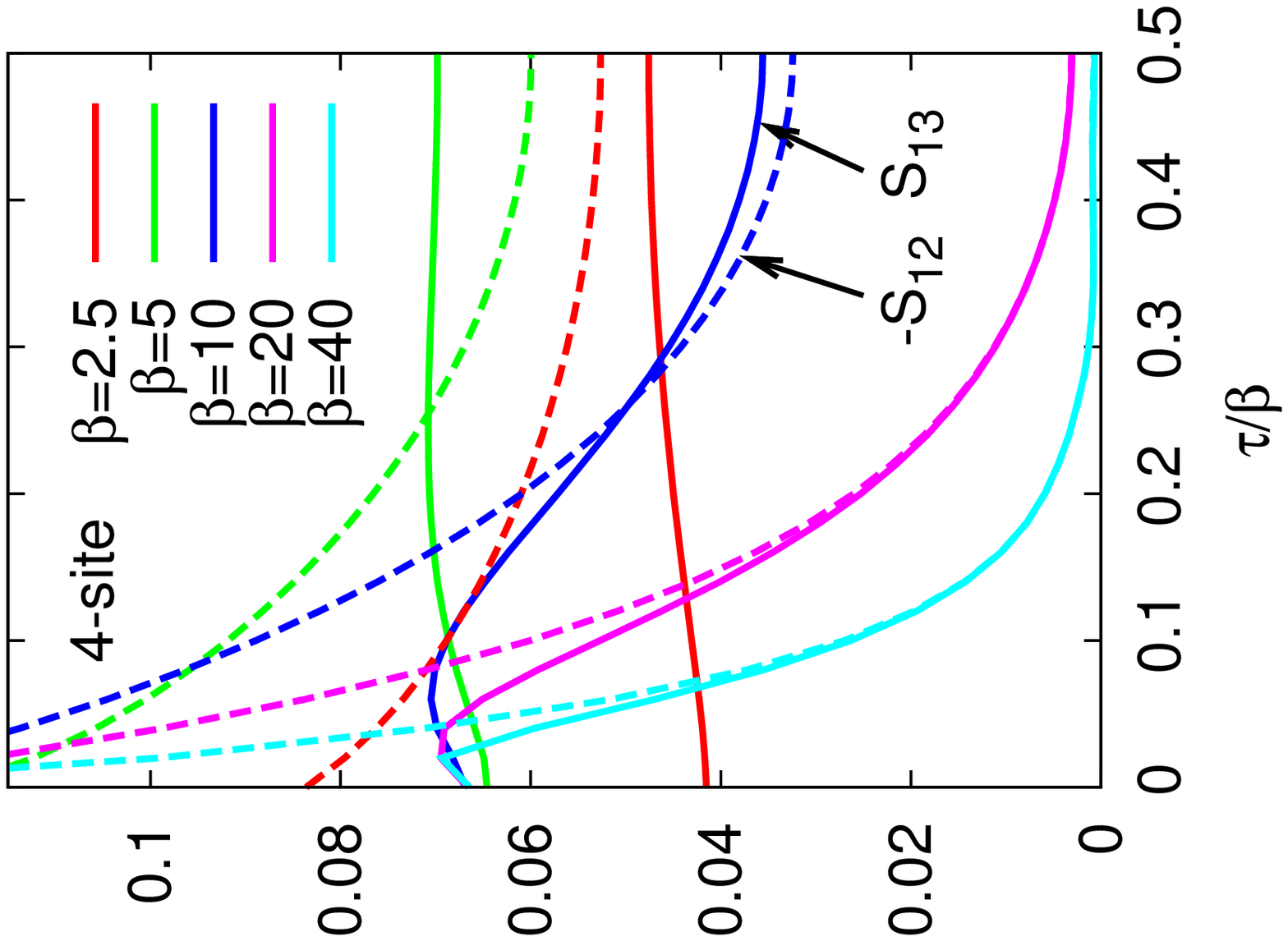}\hfill
\includegraphics[angle=-90, width=0.49\columnwidth]{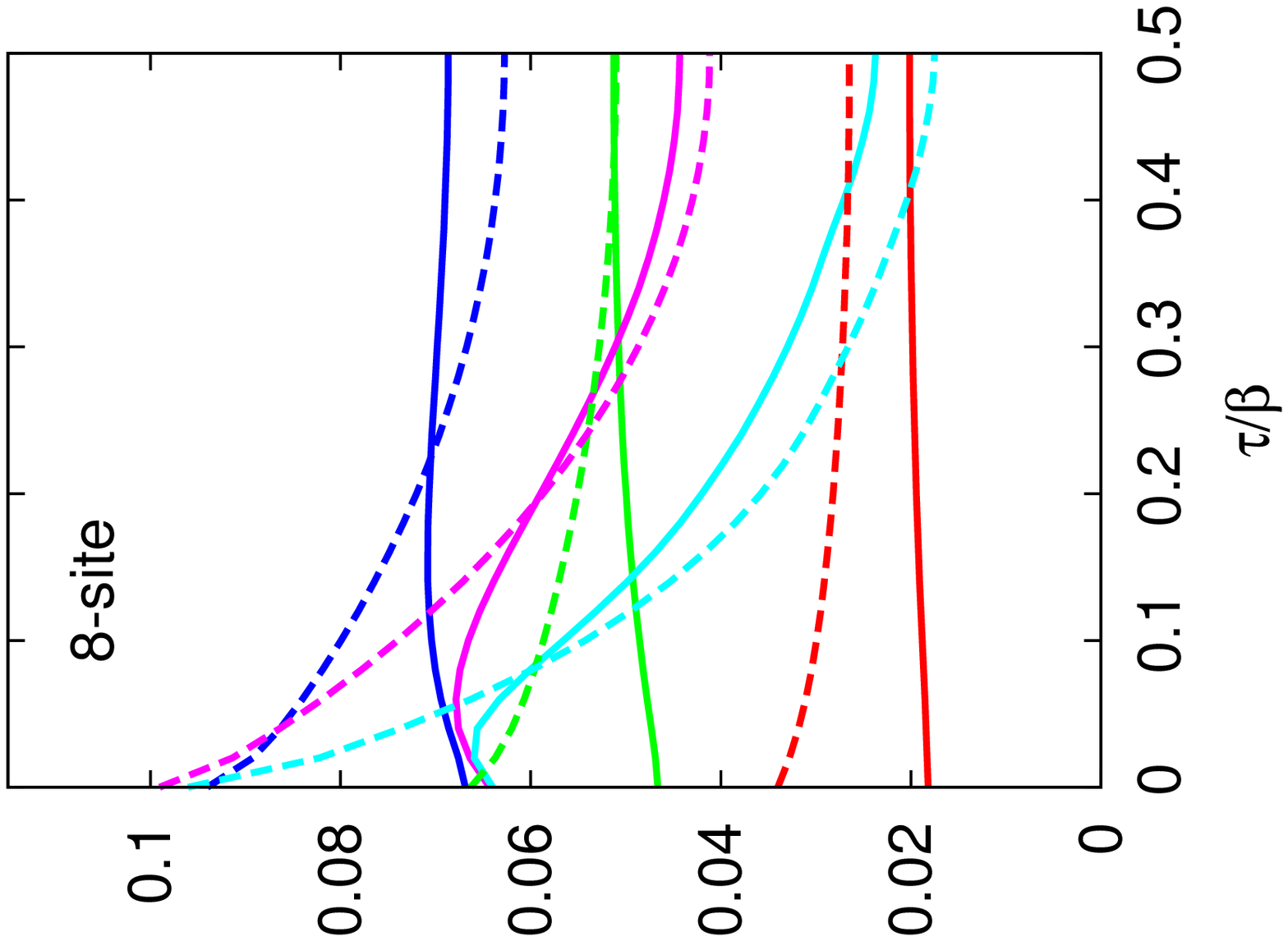}
\caption{
Imaginary-time dependent next-nearest-neighbor ($S_{13}$, solid) and nearest-neighbor ($-S_{12}$, dashed) spin correlations at half-filling and indicated $\beta$, in 4-site DCA (left) and 8-site DCA (right).}
\label{fig_halffilling}
\end{center}
\end{figure}

To investigate the apparent connection between spin-freezing and the pseudo-gap phase more closely, we have analyzed the spectral functions of the $(0,\pi)$ momentum patch, using the maximum entropy (MaxEnt) method \cite{Bryan1990,Jarrell1996,Boehnke,maxent2017}. In Fig.~\ref{fig_corrtimesbeta}, a small black dot indicates a small pseudo-gap feature, and the larger black circles a robust pseudo-gap in the anti-nodal region. While our procedure is not immune to Monte Carlo noise and uncertainties in the analytical continuation, it produces an interesting picture in rough agreement with the $T^*$ line reported in the previous literature \cite{Jarrell2000}. In particular, we find that the pseudogap region determined by MaxEnt indeed essentially coincides with the region of enhanced ferromagnetic correlations. 

The apparent weakening of the ferromagnetic tendency at low temperature and close to half-filling is due to the exponential decay of both $S_{13}$ and $S_{12}$ in this region of the phase diagram (the data for $S_{13}$ are shown in the Supplementary Material). This results in much smaller values of $\beta[S_{13}(\beta/2)-(-S_{12}(\beta/2))]$. To illustrate the crossover to an exponential decay, we plot in Fig.~\ref{fig_halffilling} the temperature dependence of $S_{13}$ and ($-S_{12}$) at half filling. The exponential decay most likely originates from strong nearest-neighbor spin-$\tfrac12$ singlet formation, see red ellipse in Fig.~\ref{fig_illustration}, which eventually leads to a break-up of the composite spin-1 moments \cite{ellipse}. It is clear though from Fig.~\ref{fig_corrtimesbeta} that this breakup happens at a lower temperature scale than the opening of the pseudogap. Hence we conclude that the opening of the pseudogap around $T^*$ is intimately connected to enhanced ferromagnetic correlations and the formation of composite spin-1 moments on the lattice diagonals. The antiferromagnetic correlations between these composite moments 
result in a pseudo-gap which is most pronounced near momentum $(0,\pi)$ and $(\pi,0)$. As temperature is lowered by a about a factor of two below $T^*$, singlet formation between nearest-neighbor spin-$\tfrac12$ becomes dominant near half filling and leads to the weakening of ferromagnetic correlations on the diagonal and an exponential decay of both $S_{12}$ and $S_{13}$. The region dominated by spin-$\tfrac12$ singlets can be associated with the downturn in $T_c$ close to half-filling, which indicates that these correlations are detrimental to superconductivity. The largest $T_c$ values are instead found in the region where the ferromagnetic $S_{13}$ correlations are strongly enhanced, while the competing spin-$\tfrac12$ singlet formation is suppressed due to doping.  

The lower panel of Fig.~\ref{fig_corrtimesbeta} shows the results obtained from 8-site DCA. The computational resources available to us do not allow to compute the $T_c$ dome, but due to the smaller mean-field effect, we expect it to peak at lower temperature. Based on the results for slightly smaller $U$ reported in Refs.~\onlinecite{Gull2013} and \onlinecite{Dong2019} we show an educated guess for the $T_c$ dome by the dashed line.  As in the 4-site DCA case, the ferromagnetic spin correlations dominate at $\tau=\beta/2$ in a wide doping region that covers the expected $T_c$ dome. The largest enhancement is found at lower temperatures in 8-site DCA, but comparable values to the 4-site case are reached for $\beta\gtrsim 10$. The suppression due to the exponential decay of the spin correlations near half-filling is barely evident at $\beta\lesssim 20$, due to the overall weaker antiferromagnetic spin correlations, see also the right panel of Fig.~\ref{fig_halffilling}. Again, we indicate by black dots and circles the filling and temperature values where we observe the opening of a pseudo-gap in the antinodal region. As in the 4-site case, the appearance of the pseudo-gap essentially coincides with the appearance of strongly enhanced $S_{13}$ correlations at long time. 

\begin{figure}[t]
\begin{center}
\includegraphics[angle=-90, width=0.9\columnwidth]{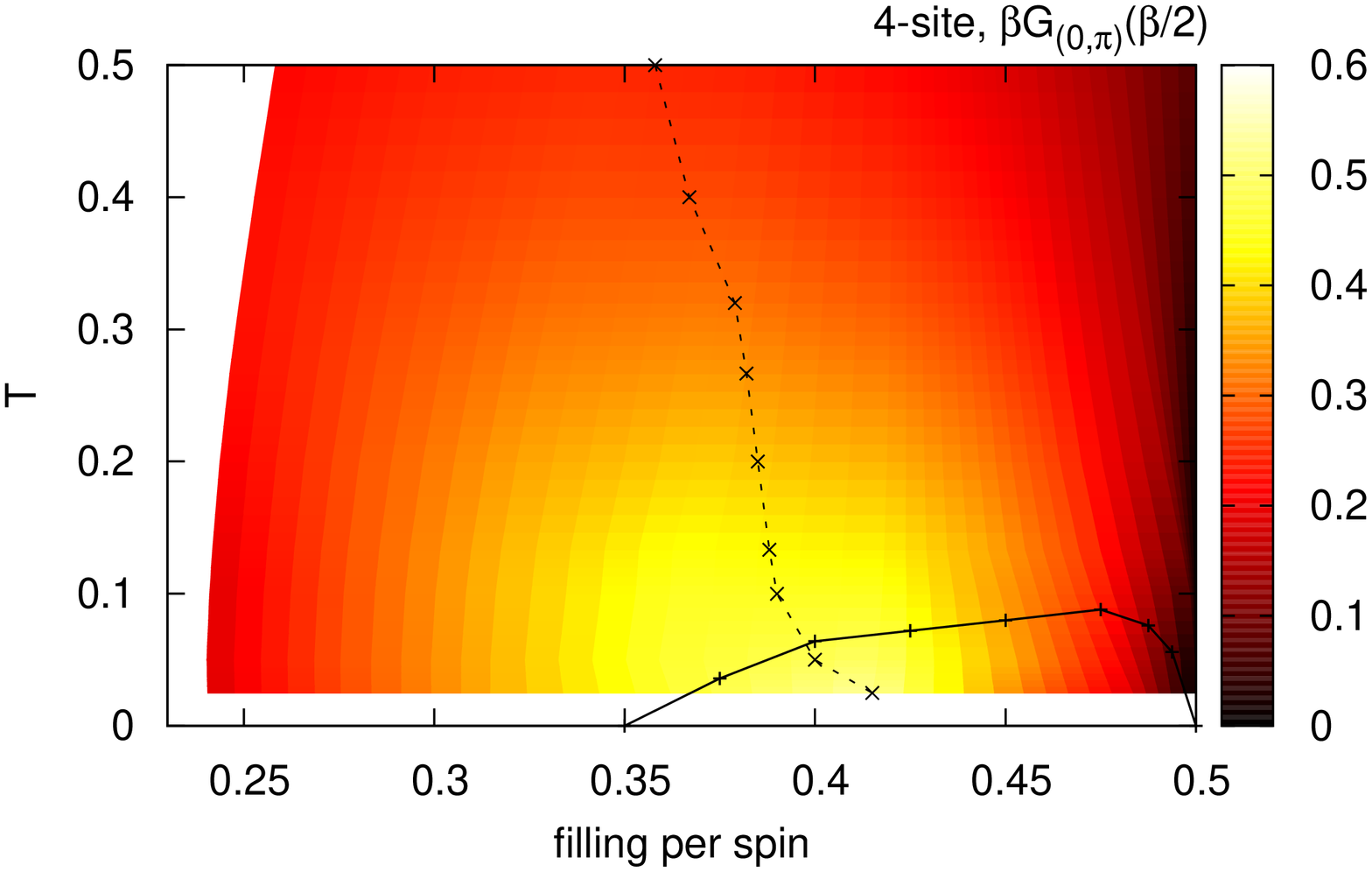}
\includegraphics[angle=-90, width=0.9\columnwidth]{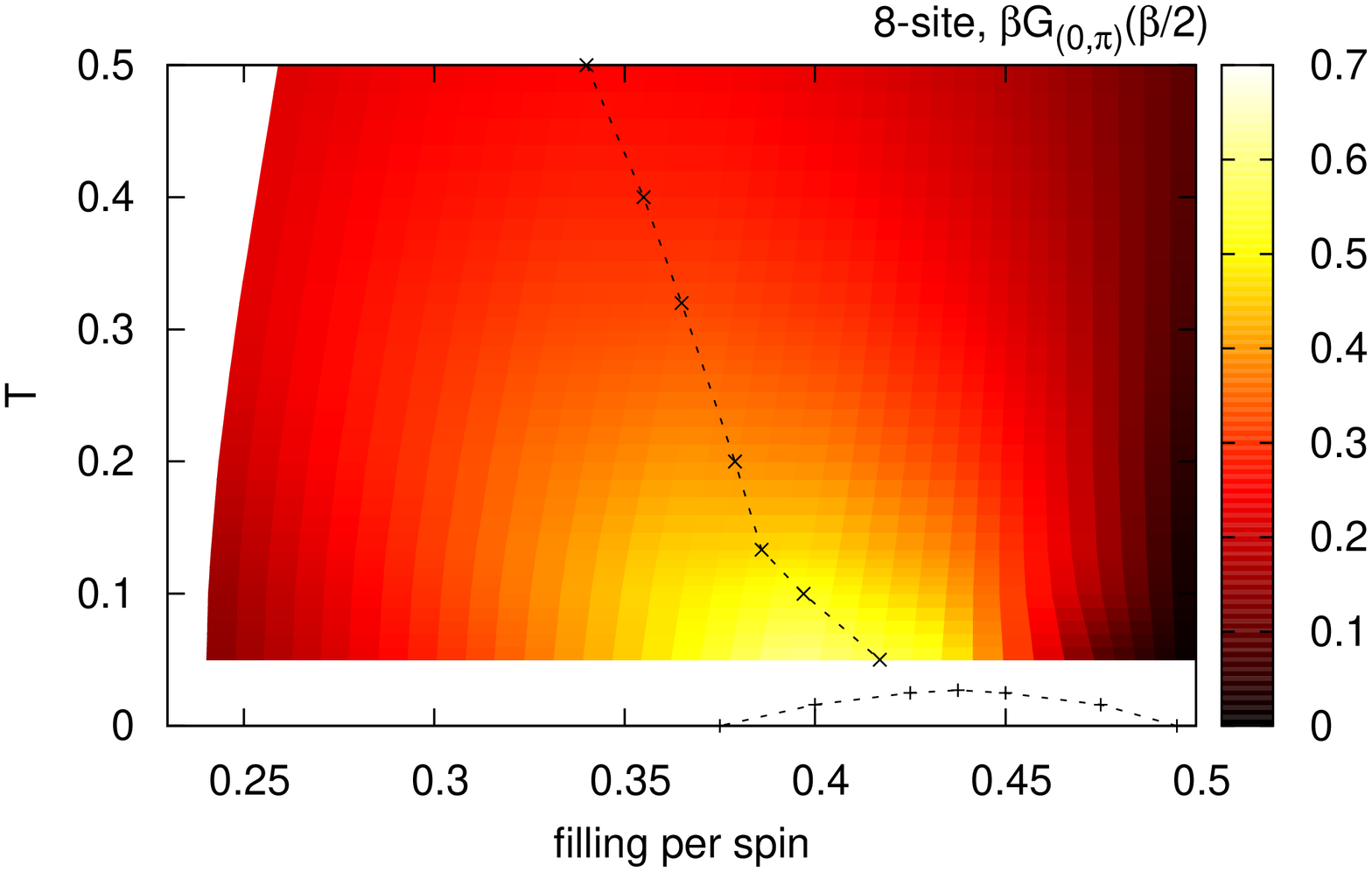}
\caption{Density of states at the Fermi level in the antinodal region, estimated from $\beta G_{(0,\pi)}(\beta/2)$. 
The dashed line shows the maximum in the broad crossover from overdoped Fermi liquid to under-doped Hund metal.  
}
\label{fig_dos_0pi}
\end{center}
\end{figure}

Apart from the pseudo-gap region, the cuprate phase diagrams feature a non-Fermi-liquid metal phase in a broad doping region at high temperatures. This behavior is expected within the spin-freezing theory and can be explained already at the level of a single-site DMFT description in the bonding/antibonding basis. This description is interesting because it disentangles the antiferromagnetic physics from the freezing of the composite spin-1 moments. As shown in Ref.~\onlinecite{Werner2016} this single-site two-orbital DMFT analysis predicts a non-Fermi-liquid crossover line which roughly extends from the half-filled $T=0$ Mott point through the peak of the superconducting dome to high temperatures and dopings. Such a crossover is also evident in Fig.~\ref{fig_dos_0pi}, which plots $\beta G_{(0,\pi)}(\beta/2)$, a quantity which is roughly proportional to the density of states at the Fermi level \cite{Gull2008}, for the Green's function corresponding to momentum patch $(0,\pi)$ (analogous plots for the local Green's function can be found in the Supplementary Material). We see that as the filling is increased from the hole overdoped region, the density of states near the Fermi level increases, reaching a maximum on the overdoped side of the superconducting dome at low temperature, and at even lower fillings at higher temperature (dashed line with diagonal crosses). Closer to half filling, the formation of local spin-1 moments and an associated Hund-metal state \cite{Werner2008,Georges2013} leads to a suppression of the density of states with increasing filling. In the analysis of the $(0,\pi)$ patch, the opening of the actual pseudo-gap is roughly visible near half-filling. A proper identification of $T^*$ however requires a direct analysis of the spectral function, as done in Fig.~\ref{fig_corrtimesbeta}, since $\beta G_{(0,\pi)}(\beta/2)$ is not sufficiently sensitive to narrow gaps. 

Finally, we mention that in the overdoped regime, the nearest-neighbor correlations (near $\tau=\beta/2$) change from antiferromagnetic to ferromagnetic.  In Fig.~\ref{fig_corrtimesbeta} we indicate the switching point of the $S_{12}$ correlations by the dashed line with diagonal crosses. It is interesting to note that several recent experiments on overdoped cuprates report ferromagnetic fluctuations \cite{Sonier2010,Kurashima2018}, which set in roughly at the doping where superconductivity disappears. The emergence of such correlations can also be qualitatively understood via the mapping to the effective two-orbital system (Fig.~\ref{fig_illustration}). In fact, the generic phase diagram of the two-orbital model with ferromagnetic Hund coupling (Fig.~3 in Ref.~\cite{Hoshino2016}) reveals that as one hole-dopes the half-filled Mott insulator, the system evolves from a region with strong antiferromagnetic correlations through a spin-freezing crossover (bad metal) region into a regime which near 3/4 filling and down to half-filling is influenced by the proximity to a ferromagnetic phase. 

In summary, we have analyzed the dynamical cluster spin correlations in the 2D Hubbard model using 4-site and 8-site DCA calculations. The diagonal next-nearest neighbor correlations exhibit a non-trivial $\tau$-dependence at $T>0$ and in the doped Mott regime, which results in robust ferromagnetic correlations at long times. This is compatible with the emergence of a composite spin-1 moment on the lattice diagonals, as a result of spin-freezing. The freezing phenomenon occurs in the doping range where superconductivity is found at low temperature, and the onset of spin-freezing approximately coincides with the opening of a pseudo-gap near the antinode. We discussed the competition between this freezing phenomenon and the formation of nearest-neighbor spin-$\tfrac12$ singlets at low temperature near half-filling. The break-up of the composite spins in favor of spin-$\tfrac12$ singlets happens at a temperature which is substantially below $T^*$, and in the underdoped region, where superconductivity is suppressed. The spin-1 moments which form at elevated temperature will freeze and order antiferromagnetically near half-filling, but they remain unfrozen down to low temperatures in the optimally doped regime. According to the spin-freezing theory of unconventional superconductivity, the amplitude fluctuations of these moments induce the attractive interaction needed for pairing \cite{Werner2016}. While we have not studied the pair susceptibility and order parameter in this work, the results of our analysis in combination with the available $T_c$ data seem to support this picture. 

{\it Acknowledgements} The calculations have been performed on the Beo04 cluster at the University of Fribourg, using a code based on ALPSCore \cite{alpscore2018}. We thank Yuan-Yao He for helpful discussions. EG was supported by NSF DMR 1606348 and PW acknowledges support from SNF Grant No. 200021\_165539. The Flatiron Institute is a division of the Simons Foundation.

\clearpage  

\widetext

\begin{center}
{\centering \bf Supplementary Material}
\end{center}

\begin{figure}[h!]
\begin{center}
\includegraphics[angle=-90, width=0.425\columnwidth]{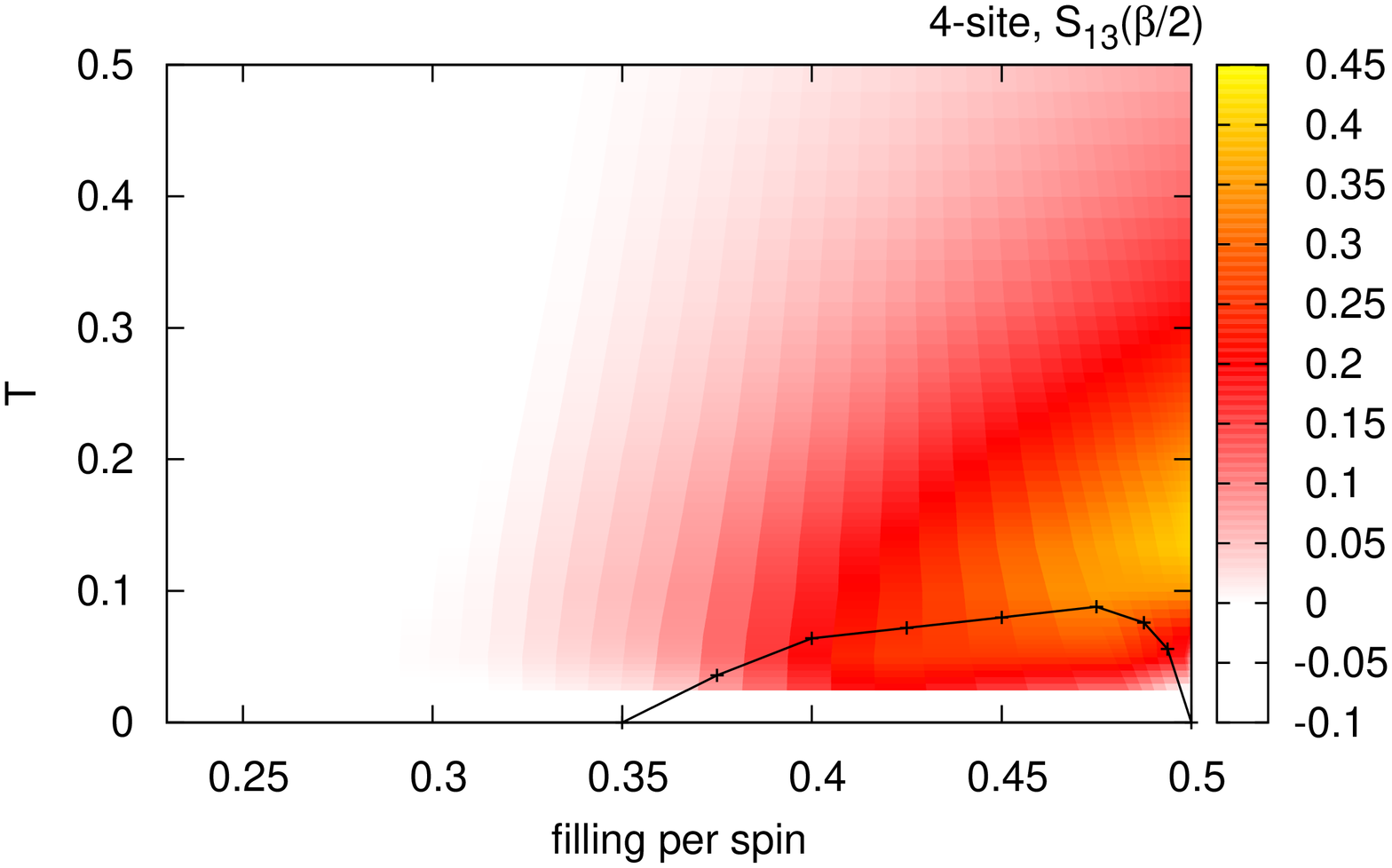}
\includegraphics[angle=-90, width=0.425\columnwidth]{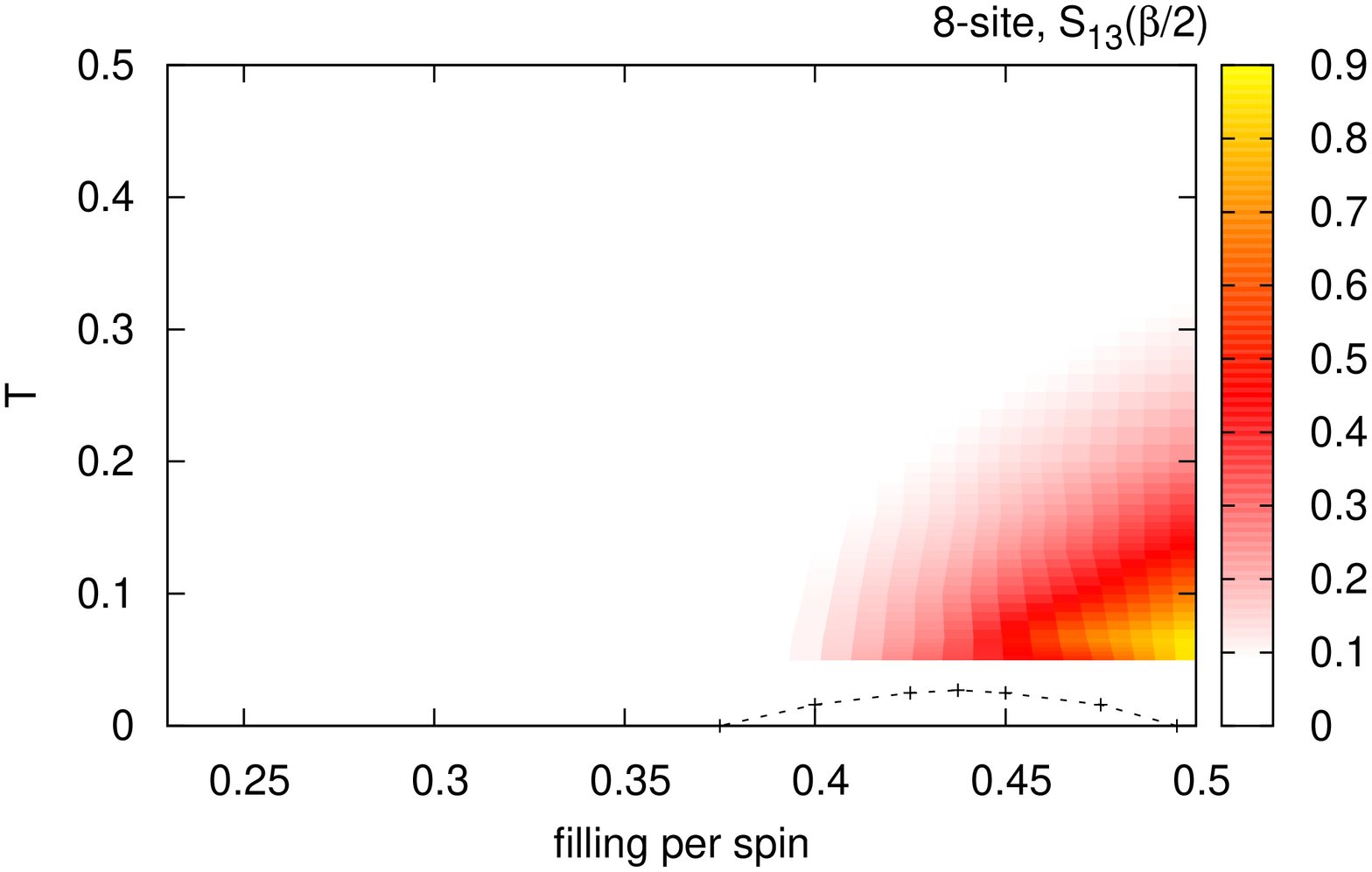}
\caption{
Long-time value of the next-nearest-neighbor spin correlation, $S_{13}(\beta/2)$, in 4-site DCA (left) and 8-site DCA (right).}
\label{fig_corr2}
\end{center}
\end{figure}

These supplementary notes provide additional data for spin correlation functions and Green functions. 
Figure~\ref{fig_corr2} presents the $S_{13}$ correlation functions at $\tau=\beta/2$. The results for $-S_{12}$ are qualitatively similar. In these plots one can recognize the appearance of the composite moments at roughly $T^*$ and the suppression due to spin-$\tfrac12$ singlet formation at low temperature near half-filling.  

\begin{figure}[h!]
\begin{center}
\includegraphics[angle=-90, width=0.425\columnwidth]{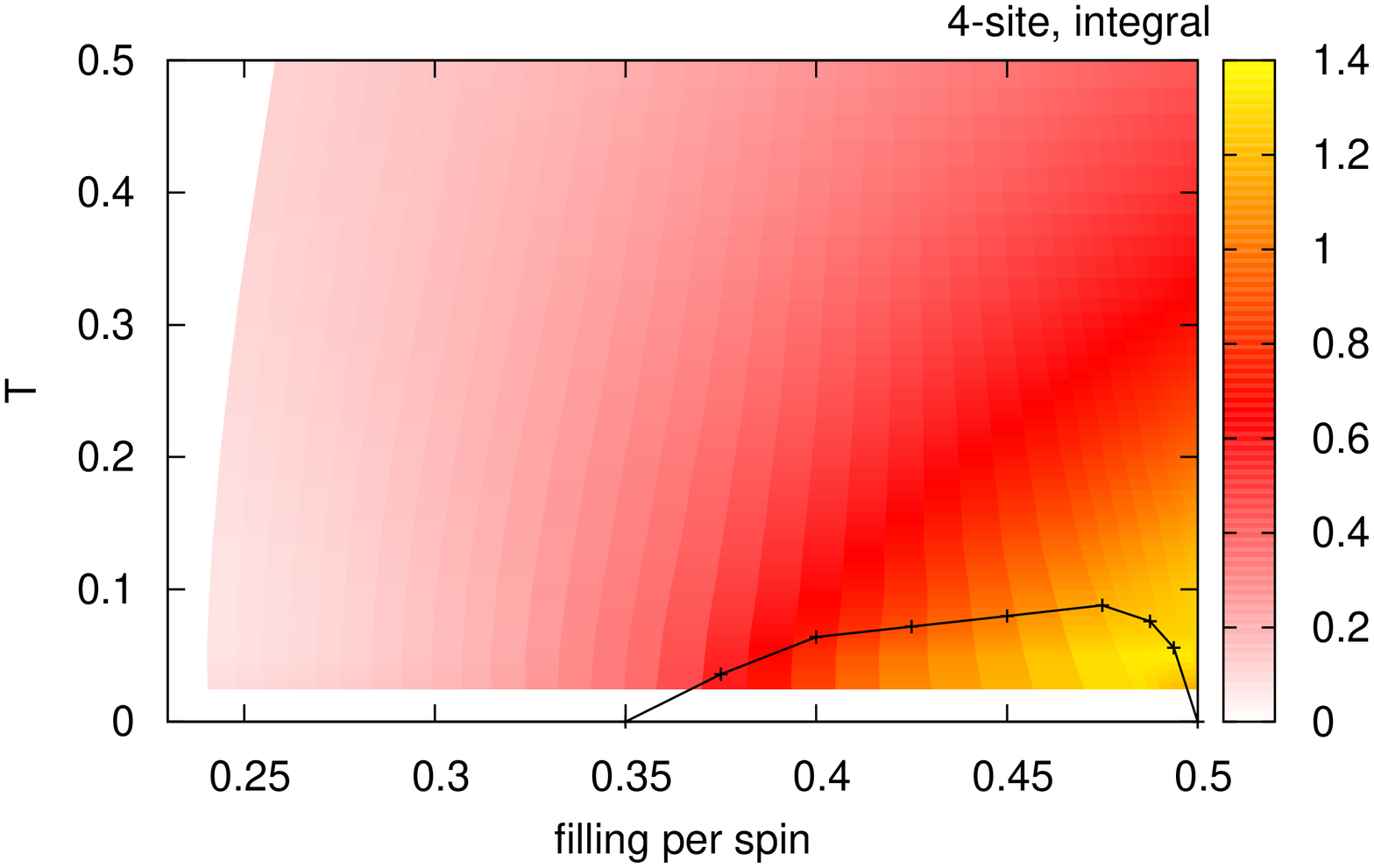}
\includegraphics[angle=-90, width=0.425\columnwidth]{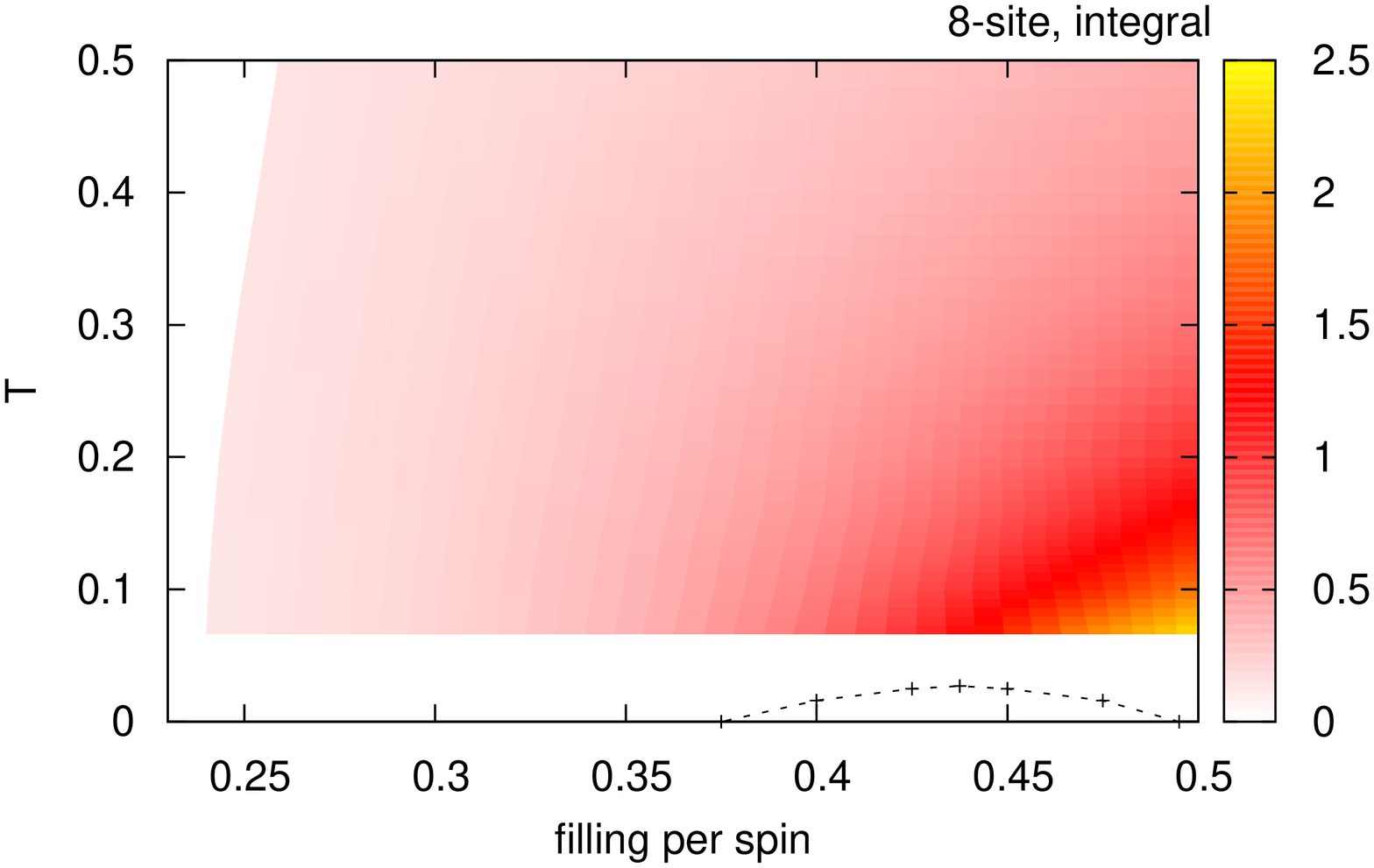}\\
\includegraphics[angle=-90, width=0.425\columnwidth]{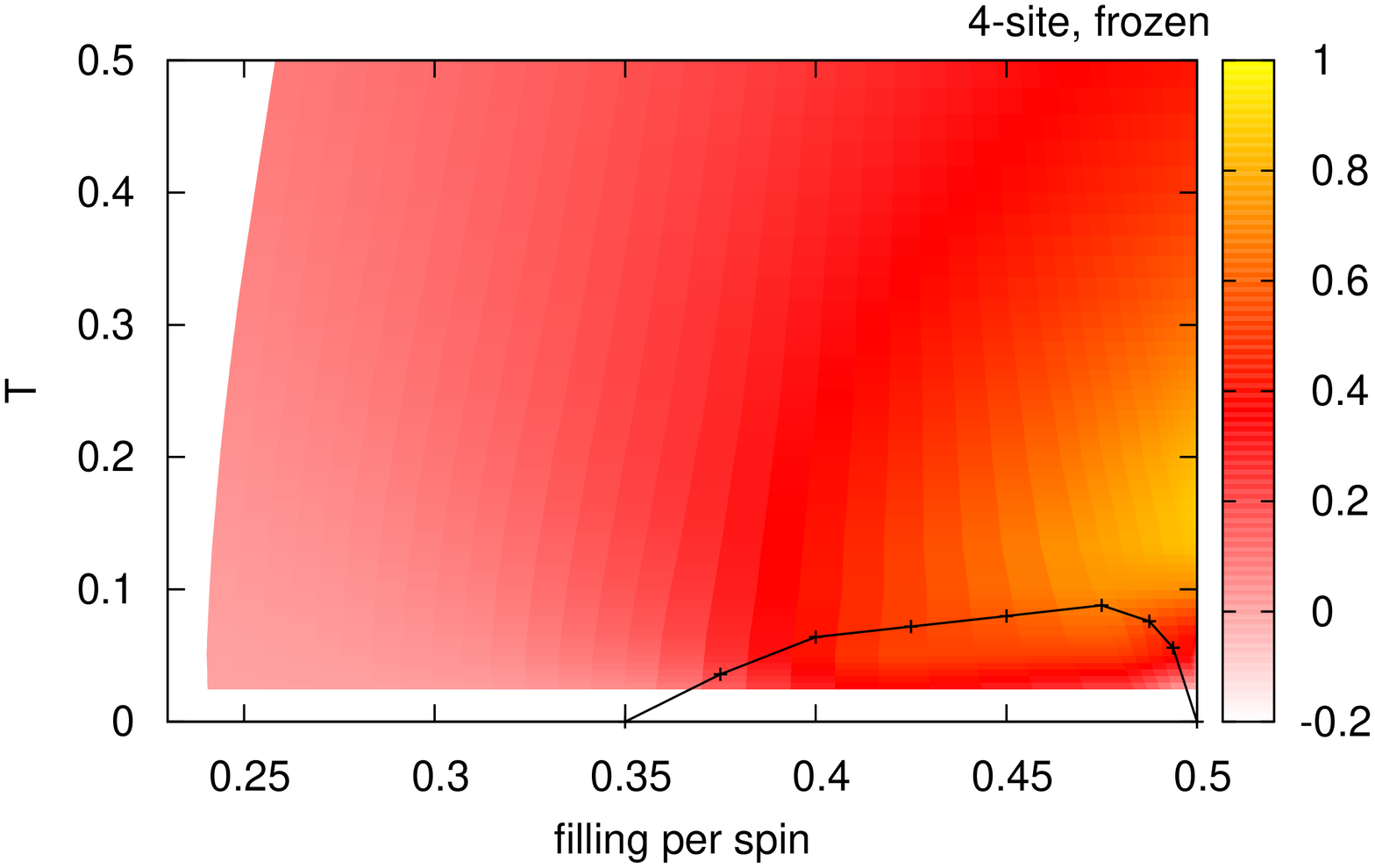}
\includegraphics[angle=-90, width=0.425\columnwidth]{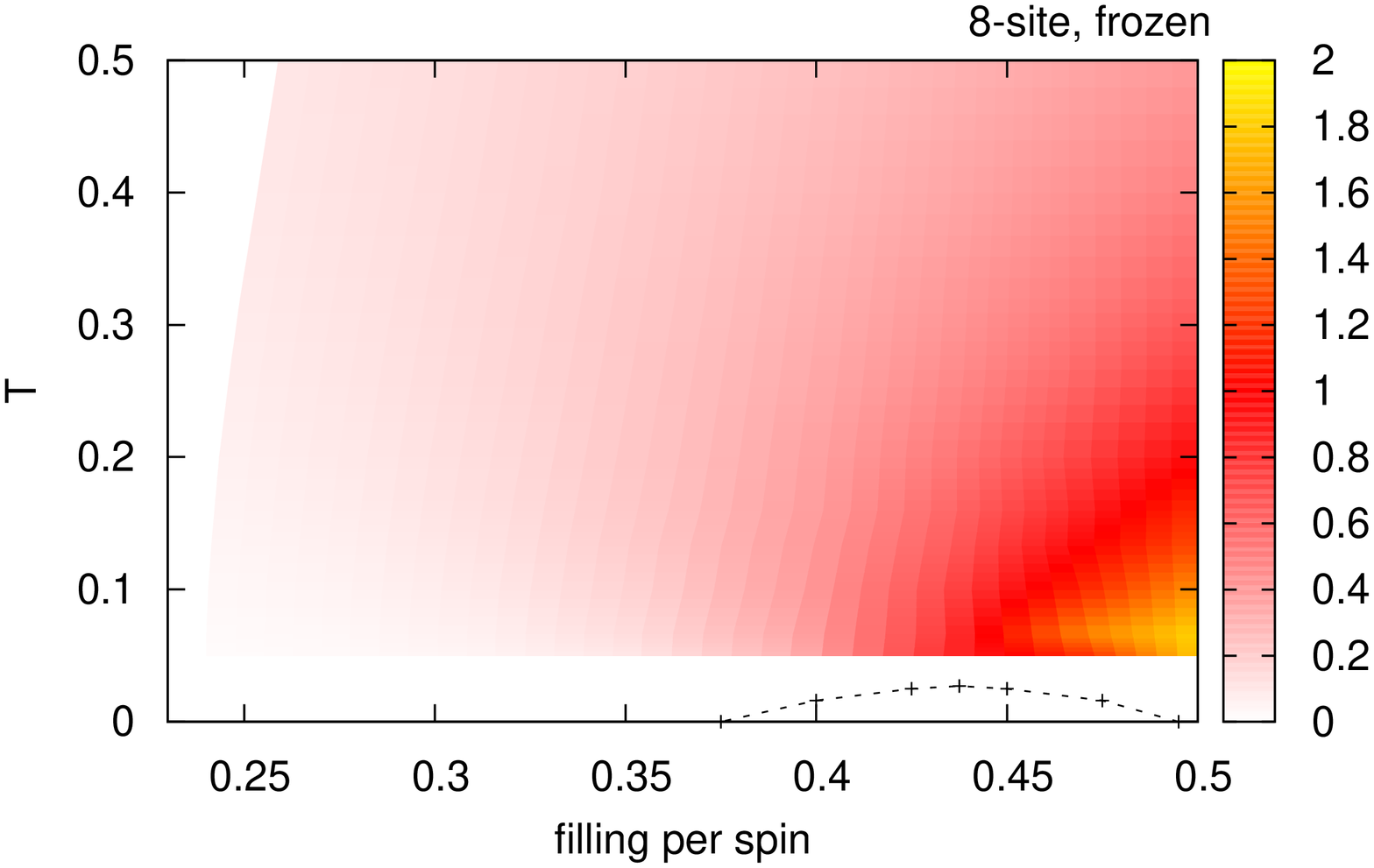}
\caption{
Total (top) and frozen (bottom) contribution to the integral of $S_{11} + S_{13}$ in 4-site DCA (left) and 8-site DCA (right).}
\label{fig_corr3}
\end{center}
\end{figure}

The top panels of Fig.~\ref{fig_corr3} show the integral of the correlation function $S_{11}+S_{13}$, $\int_0^\beta d\tau (S_{11}(\tau)+S_{13}(\tau))$, which is the easily measurable contribution to the $S_{ff}$ correlation function. The bottom panels report the contribution to this integral from frozen moments, $\beta[S_{11}(\beta/2)+S_{13}(\beta/2)]$.

\begin{figure}[t]
\begin{center}
\includegraphics[angle=-90, width=0.425\columnwidth]{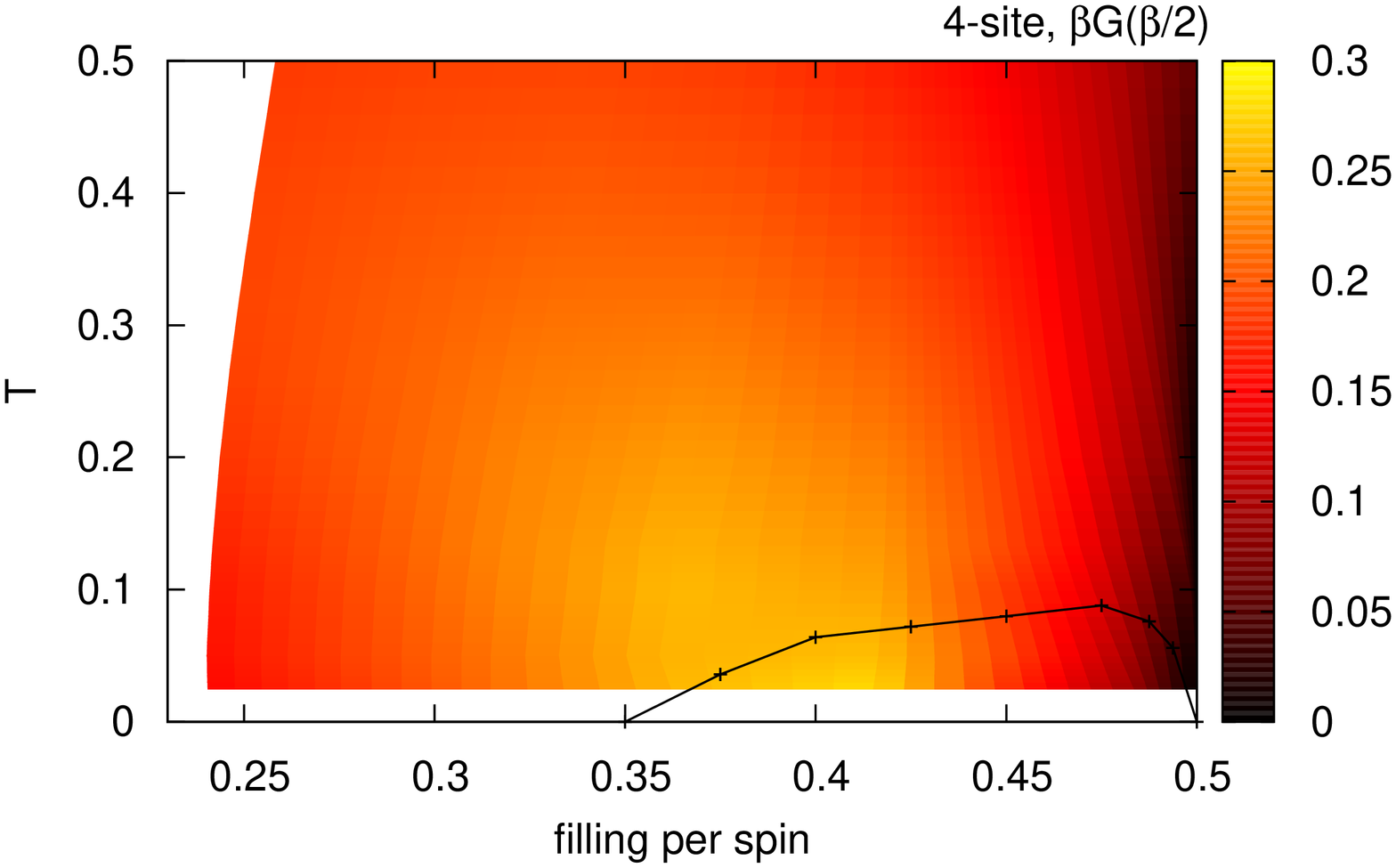}
\includegraphics[angle=-90, width=0.425\columnwidth]{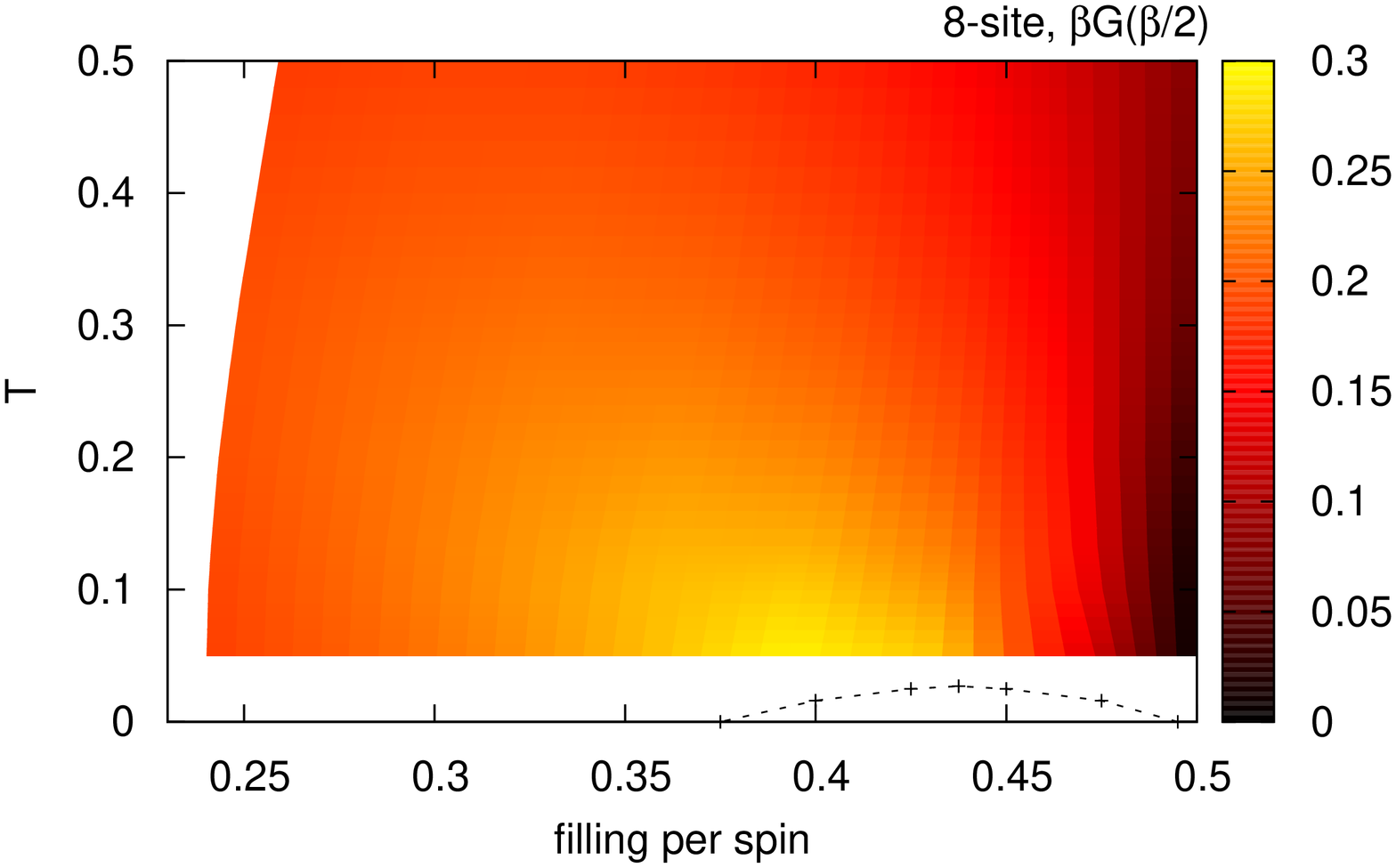}
\caption{Density of states at the Fermi level, estimated from the local Green's function $G$ as $\beta G(\beta/2)$.  
}
\label{fig_dos}
\end{center}
\end{figure}

Finally, we plot in Fig.~\ref{fig_dos} the density of states at the Fermi level, $\beta G(\beta/2)$, evaluated with the local Green's function.  Similar to Fig.~\ref{fig_dos_0pi} (results for the $(0,\pi)$ patch), these data illustrate the crossover from a Fermi liquid like metal with low carrier density in the overdoped regime to an incoherent Hund metal with suppressed density of states in the underdoped regime. 

\end{document}